\documentclass[%
aip,
amsmath,amssymb,
reprint,%
]{revtex4-1}

\usepackage{graphicx}
\usepackage{dcolumn}
\usepackage{bm}

\usepackage[utf8]{inputenc}
\usepackage[T1]{fontenc}
\usepackage{mathptmx}
\usepackage{etoolbox}
\usepackage[normalem]{ulem}
\usepackage{subcaption}
 \usepackage[version=4]{mhchem}

\makeatletter
\def\@email#1#2{%
	\endgroup
	\patchcmd{\titleblock@produce}
	{\frontmatter@RRAPformat}
	{\frontmatter@RRAPformat{\produce@RRAP{*#1\href{mailto:#2}{#2}}}\frontmatter@RRAPformat}
	{}{}
}%
\makeatother
\begin{document}
	
	\preprint{AIP/123-QED}
	
	\title{Understanding Radicals via Orbital Parities}
	\author{Reza G. Shirazi}
	
	\author{Benedikt M. Schoenauer}
	
	\author{Peter Schmitteckert}
	
	\author{Michael Marthaler}

	\author{Vladimir V. Rybkin}
	\email{vladimir.rybkin@quantumsimulations.de}
	\affiliation{$^1$HQS Quantum Simulations GmbH, Rintheimer Str. 23, 76131 Karlsruhe, Germany}


\begin{abstract}
We introduce analysis of orbital parities as a concept and a tool for understanding radicals. Based on fundamental reduced one- and two-electron density matrices, our approach allows us to evaluate a total measure of radical character and provides spin-like orbitals to visualize real excess spin or odd electron distribution of singlet polyradicals. Finding spin-like orbitals aumotically results in their localization in the case of disjoint (zwitterionic) radicals and so enables radical classification based on spin-site separability. We demonstrate capabilities of the parity analysis by applying it to a number of polyradicals and to prototypical covalent bond breaking.
\end{abstract}

\maketitle

\section*{Introduction}
\label{introduction}
Importance of radicals for chemistry cannot be overestimated due to their role as reactive intermediates\cite{zard2003radical}, magnetic\cite{de2015magnetic} and optical properties\cite{Nakano_Champagne}, and biological function\cite{radicals_bio}. Theoretical chemistry has achieved impressive results in understanding their structure and reactivity as well as in accurate calculation of their properties\cite{Nakano2017, Krylov_book}. At the same time, the general picture is still surprisingly non-uniform, leading to several widely used classifications\cite{Hoffmann_chem_rev, diradicals_Salem}.

The conventional definition of a radical as a system with unpaired electron(s) is intuitive, but barely applicable in cases where the electronic structure is not described by a simple valence-bond picture. On the other hand, a physical view based on quantum numbers of the electronic wave function, the total spin $S^2$ and its projection $m_s$, is not insightful for some systems. An iconic example of those are singlet diradicals\cite{Hoffmann_chem_rev, diradicals_Salem, diradicals_Abe, Nakano2017}. For them both quantum numbers are zero, suggesting a simple singlet molecule. Neither is spin density insightful, as it is zero everywhere\cite{Staroverov2000}. However, spectroscopic and chemical properties of these molecules reveal radical behavior. 

An ideal theoretical framework for understanding radicals should provide a numerical measure for radical (polyradical) character of the molecule and a function for the unpaired electron distribution suitable for visualization. In addition, the insights into the ability of multiple radical sites to react independently would be a bonus as it helps classifying polyradicals into disjoint and non-disjoint\cite{Hoffmann_chem_rev, diradicals_Abe}, also known as zwitterionic and covalent \cite{diradicals_Salem}, types. 


There exist numerous valuable theoretical approaches for understanding radicals based on various properties: natural occupation numbers \cite{Yamaguchi1975TheES}, configuration interaction coefficients \cite{Bachler, Hayes}, collectivity number \cite{Luzanov2005}, hole-particle density \cite{Luzanov_Prezhdo}, distribution of unpaired electrons \cite{Takatsuka1978, Staroverov2000, HEADGORDON2003508}, natural \cite{Doehnert} or partial orbital occupations \cite{FOD}. Many of these theories are conceptually involved and often based on specific or complicated wave-function types.

Here we propose an alternative framework for understanding radicals based on orbital parity, inspired by the theory of Mott insulators ("mottness")\cite{Mottness}. Our approach is conceptually simple, not restricted to particular types or systems and wave functions, and offers a qualitative measure of the radical character. Moreover, our method provides a transformation to a spin-like orbital basis suitable for visualizing the unpaired electron distribution. This basis may or may not be localized, thus, making it possible to distinguish between disjoint and non-disjoint polyradicals. In addition, spin-like orbitals can be used to simplify post-processing quantum chemical calculations. 
There already exist successful theoretical approaches to radical chemistry based on orbital transformations\cite{Amos_Hall, Pulay_Hamilton, NEESE2004781}. However, we argue that using the transformation based on orbital parity optimization demonstrates a number of lucrative features.

\section*{Results and Discussion}
The energy of the molecule in the normalized electronic state $|0\rangle$ is given as \cite{helgaker_bible}
\begin{equation} \label {Hamiltonian}
E = \sum_{p, q} D_{pq} h_{pq} + \frac{1}{2} \sum_{pqrs} d_{pqrs} g_{pqrs} + h_{nuc},
\end{equation}\\
where indices ${p,q,r,s}$ run over spatial orbitals; $D_{pq}$ and $d_{pqrs}$ are one-electron and two electron reduced density matrix (1-RDM and 2-RDM) elements in the molecular orbital (MO) basis, respectively; $h_{pq}$ and  $g_{pqrs}$ are one-electron integrals and two electron integrals, respectively, and $h_{nuc}$ stands for nuclei Coulomb repulsion. RDMs are defined as follows:
\begin{eqnarray}
D_{pq} = \sum_{\sigma} \langle 0 |\hat{a}_{p \sigma}^{\dag} \hat{a}_{q \sigma} | 0 \rangle, \\
d_{pqrs} = \sum_{\sigma {\sigma}^{'}} \langle 0 | \hat{a}_{p \sigma}^{\dag} \hat{a}_{r {\sigma}^{'}}^{\dag} \hat{a}_{s {\sigma}^{'}} \hat{a}_{q {\sigma}} | 0 \rangle
\end{eqnarray}
In these equations $\sigma$, $\sigma^{'}$ correspond to the spin functions and run over spin-up, $\uparrow$, and spin-down, $\downarrow$, states, whereas $a_{p \sigma}^{\dag}$ and $a_{p \sigma}$ are standard second-quantization creation and annihilation operators. 
%
%

We now define the parity operator for orbital $p$ using the number operators $\hat{n}_{p\uparrow} = \hat{a}_{p \uparrow}^{\dag} \hat{a}_{p \uparrow}$ and $\hat{n}_{p\downarrow} = \hat{a}_{p \downarrow}^{\dag} \hat{a}_{p \downarrow}$ (expectation values of these, ${n}_{p\uparrow}$ and ${n}_{p\downarrow}$, being spin-orbital occupations):
\begin{equation} \label{parity_ON}
\hat{P}_p = (-1)^{\hat{n}_{p\uparrow} + \hat{n}_{p\downarrow}}.
\end{equation}
The parity expectation value is calculated from RDMs:
\begin{equation} \label{parity_DM}
P_p = \langle 0 | \hat{P}_p | 0 \rangle = 1 - 2 D_{pp} + 2d_{pppp}.
\end{equation}
$n_p\uparrow$ and $n_p\downarrow$ take values between 0 and 1. For single determinant methods, the spin-orbital occupations are exactly 0 (empty) or 1 (occupied), whereas for more general, multiconfigurational, wave functions they are non-integer. Orbital occupation is the sum of $n_p\uparrow$ and $n_p\downarrow$.
From equation \eqref{parity_ON} we see that $P_p$  takes values from -1 to 1. For the single-determinant wave function singly-occupied orbitals have $P_p = -1$, whereas for vacant or doubly-occupied ones $P_p = 1$ as listed in Table \ref{tab:occ_par}. Thus, there is a clear mapping between orbital occupation number and its parity: half-filled orbitals are spin-like.

\begin{table}
	\begin{center}
		\caption{Relation between spin-orbital occupation numbers and orbital parities for a single Slater determinant.}\label{tab:occ_par}
		\begin{tabular}{ccr}	
			\toprule		
			$n_{p\uparrow}$ & $n_{p\downarrow}$ & $P_p$ \\
			\hline
			0	&	 0	&	1	\\
			1 	&	 0	&	-1	\\
			0   &	 1  &	-1 \\
			1 	&	 1	&	1	\\
			\hline
		\end{tabular}
	\end{center}
\end{table}

In the case of multiconfigurational wave funcitons, the simple mapping between occupations and parities does not generally hold. Singly occupied molecular orbitals (canonical, natural \textit{etc.}) may not possess spin-like character. However, it may be possible to minimize the parity, bringing it close to -1 with an orbital transformation. This can be achieved by pairwise orbital rotations by an angle $\phi$ and satisfying the minimum conditions:
\begin{equation}
\label{eq:cond_rot}
\frac{d P_p (\phi)}{d\phi} = 0\,,
\end{equation}
and selecting the solutions corresponding to the parity minimum.
%
where $P_p (\phi)$ is the orbital parity in the transformed molecular orbital basis. Therefore, within our framework \textbf{a radical is a system where one or more orbitals with $P_p = -1$ can be found} by orbital rotations. We will refer to such orbitals as \textbf{spin-like}. Diradicals have two spin-like orbitals, triradicals have three \textit{etc.}  In trivial cases such as those in Table \ref{tab:occ_par}, the transformation from the canonical basis is a unit matrix. 

In case of polyradicals, parity optimization may result in orbital localization. \textbf{Spatial distribution of the spin-like orbitals provides a tool to classify radicals into the disjoint and non-disjoint classes}: in the former, the spin-like orbitals are localized on different sites. 

For the general multiconfigurational wave functions required to capture electron correlation, \textbf{parities are not integer (see equation \eqref{parity_DM})and can naturally serve as measures of radical character}. As opposed to some criteria introduced for diradicals $ P_p(\phi)$ are a general criterion, \textit{i. e.} applicable to all radical types. 

RDMs, available from standard quantum chemistry software, are needed to perform the spin-mapper orbital transformation that searches for spin-like orbitals. The latter has been derived and implemented by us and is described in more detail in the Appendix. 

As a first example, we apply the orbital parity formalism to singlet diradicals{\cite{Hoffmann_chem_rev, diradicals_Abe, diradicals_Salem}. We consider benzyne isomers a pyridine-based analogue of p-benzyne \cite{polar} shown in Figure \ref{fig:formulae}. Orbital parity analysis of the CASSCF(8,8) \cite{Lindh_cas} wave functions of o- and m-benzynes reveals a weak diradical character with minimum parities of -0.5 and -0.3, respectively (see Figure \ref{fig:parities}). This is to be expected due to close proximity of the supposed unpaired electrons, which are able to couple with each other contributing to a triple bond\cite{Shaik_benzynes}. In contrast to this, p-benzyne and its analogue are proper diradicals with two spin-like orbitals each ($P\approx$ -0.9).

	%
	%
	\begin{figure}[ht]
		\centering
		\begin{subfigure}[t]{.3\linewidth}
			\includegraphics[width=.55\linewidth]{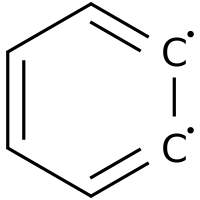}
			\caption*{\textbf{o-b}}\label{fig:pb_form}
		\end{subfigure}
		\begin{subfigure}[t]{.3\linewidth}
			\includegraphics[width=.55\linewidth]{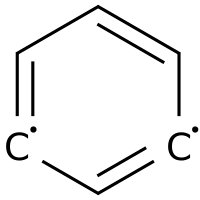}
			\caption*{\textbf{m-b}}\label{fig:mb_form}
		\end{subfigure}
		\begin{subfigure}[t]{.3\linewidth}
			\includegraphics[width=.55\linewidth]{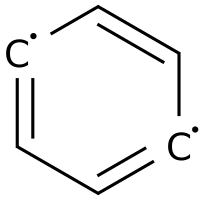}
			\caption*{\textbf{p-b}}\label{fig:pb_form}
		\end{subfigure}

		\begin{subfigure}[t]{.3\linewidth}
			\includegraphics[width=.6\linewidth]{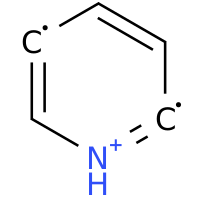}
			\caption*{\textbf{ddp-1} }\label{fig:polar-1}
		\end{subfigure}
		\begin{subfigure}[t]{.3\linewidth}
			\includegraphics[width=0.9\linewidth]{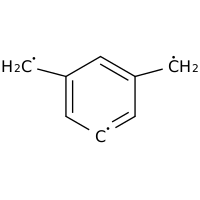}
			\caption*{\textbf{dmx} }\label{fig:dmx}
		\end{subfigure}
		\caption{Chemical structures of the considered radicals: \textbf{o-b} - ortho-benzyne; \textbf{m-b} - meta-benzyne; \textbf{p-b} - para-benzyne;
			\textbf{ddp-1} - 2,5-didehydropyridinium cation; \textbf{dmx} - 5-dehydro-m-xylylene.}
		\label{fig:formulae}
	\end{figure}
	
	For p-benzyne, parity optimization results in the spin-like orbitals being a symmetric and antisymmetric combination of the canonical HOMO and LUMO. This is to be expectred for the multi-configurational wave function dominated by the two configurations: doubly-occupied HOMO with empty LUMO and \textit{vice versa}. Whereas canonical orbitals are delocalized, spin-like ones are localized at different carbon atom as shown in Figure \ref{fig:p-benzyne} indicating that p-benzyne is a disjoint (zwitterionic) diradical. This transformation (45$^{\circ}$ rotation) is a well-known result of generalized valence-bond (GVB) theory \cite{GVB, Hoffmann_chem_rev, unusual_biradicals}. Although the canonical HOMO and LUMO in p-benzyne are have occupation numbers close to one, their parities not spin-like (see Figure \ref{fig:p-benzyne}).
	
	%
	%
	
	\begin{figure}[ht]
		\begin{center}
			\includegraphics[width=8.cm]{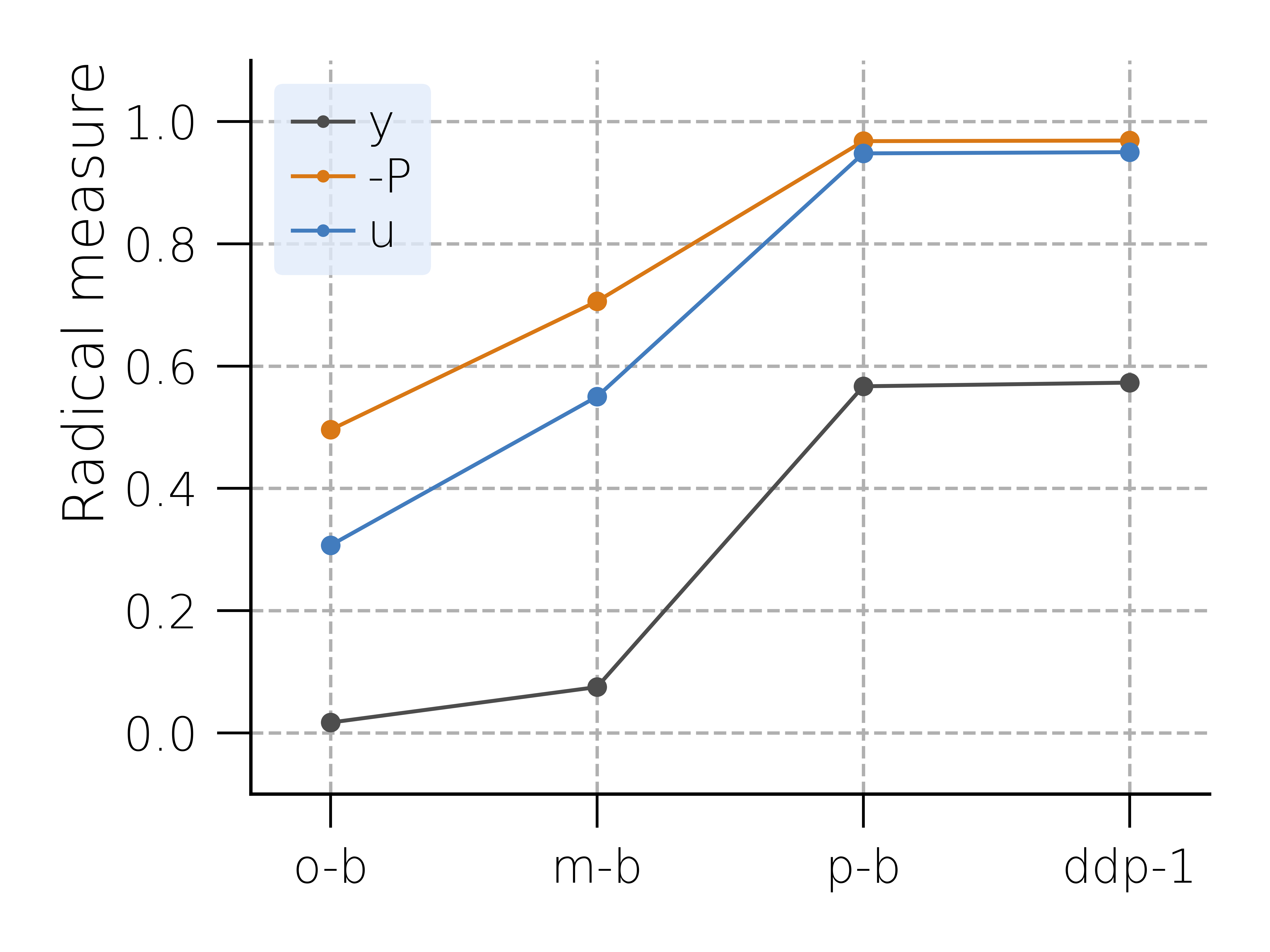}
			\caption{Radical measures (parities, odd electron distribution functions, and diradical characters) for benzynes and analogues: $-P$ -- negative parity; $u$ -- odd electron density; $y$ -- Yamaguchi's diradical character. $-P$ is the average over the two spin-like orbitals (are identical due to symmetry). $u$ is averaged over the two canonical frontier orbitals.}
			\label{fig:parities}
		\end{center}
	\end{figure}
	
	Pyridine-based diradical is less symmetric than p-benzyne, which results in a more general spin-mapper transformation: rotation angle different from 45$^{\circ}$. GVB-tranformed orbitals are similar to the spin-like ones, but exhibit non-optimal parities (see Figure \ref{fig:polar_1}). 
	
	As shown in Figure \ref{fig:parities}, parity analysis generally agrees with the odd electron distribution \cite{Takatsuka1978, Staroverov2000, HEADGORDON2003508} results for overall diradical character, whereas Yamaguchi's function \cite{Yamaguchi1975TheES} underestimates it. Importantly, the odd electron distribution function $u(\mathbf{r})$ (see the Appendix) shows similar trends as the two spin-like orbitals, but does not provide information on the separability of the radical sites.

	%
	%
	\begin{figure}[ht]
		\begin{subfigure}[b]{8.3cm}
			\centering
			\includegraphics[width=8.3cm]{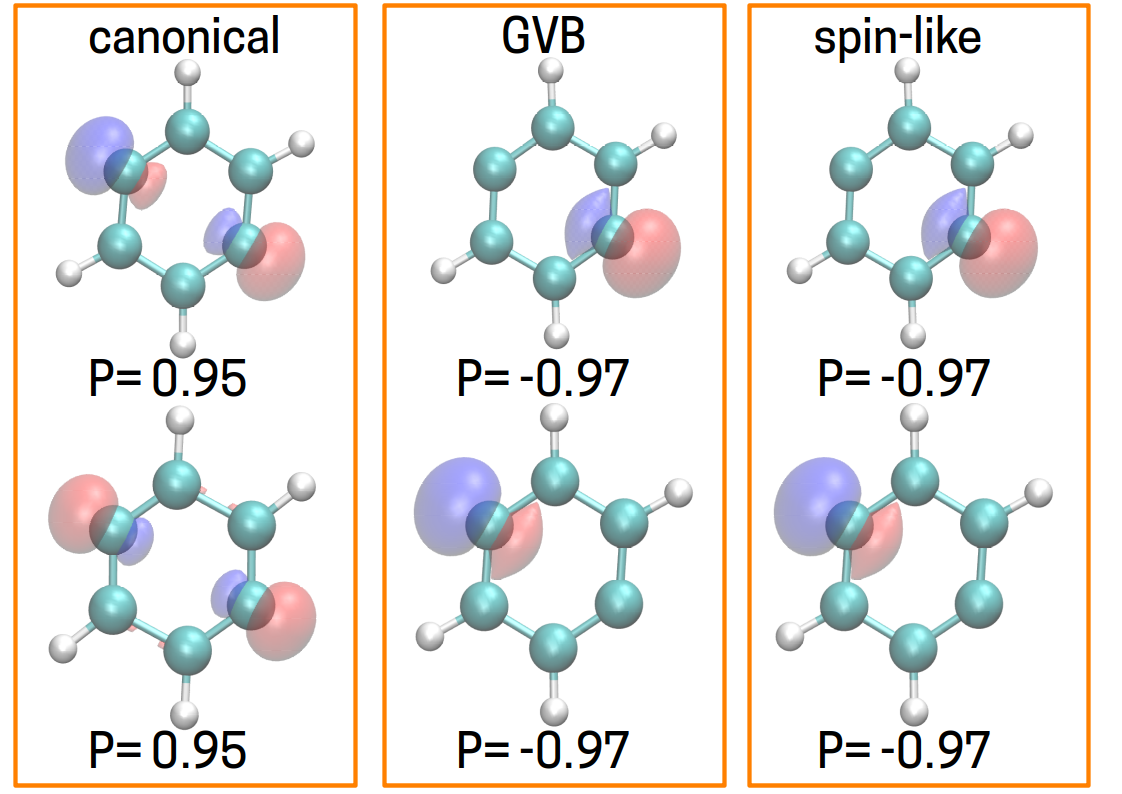}
			\caption{p-benzyne (p-b)}
			\label{fig:p-benzyne}
		\end{subfigure}
		\vfill
		\begin{subfigure}[b]{8.3cm}
			\includegraphics[width=8.3cm]{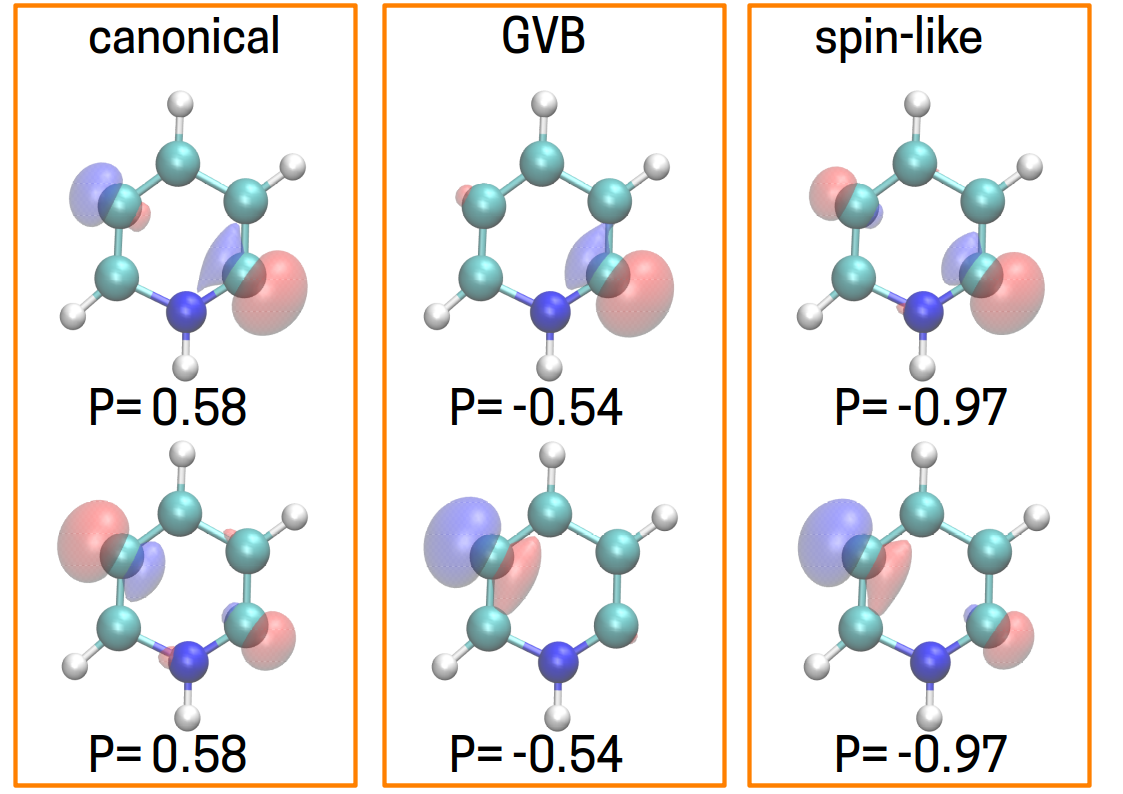}
			\caption{2,5-didehydropyridimium cation (ddp-1)}
			\label{fig:polar_1}
		\end{subfigure}
		\caption{Molecular orbitals of diradicals and their parities: canonical, generalized-valence-bond (GVB) transformed, and spin-like.}
		\label{fig:diradicals}
	\end{figure}
	%
	%
	
	We go on by applying the parity-based approach to a triradical, 5-dehydro-m-xylylene, or \textbf{dmx} (see Figure \ref{fig:formulae}) \cite{triradicals_rev}. This remarkable molecule was the first organic compound to known to violate Hund's rule as it has an open-shell doublet rather than quartet ground state\cite{slipchenko_dmx}. Our state-averaged CASSCF(9,9) calculation confirms this, revealing three almost perfectly spin-like orbitals in the ground state shown in Figure \ref{fig:dmx_sm}. The $sp^2$-hybrid one localized on the benzene ring (rightmost) has the unpaired electron on it, and is identical to the canonical orbital (\ref{fig:dmx_canon}) with perfect parity of -1. The other two spin-like orbitals localize (although not completely) on the different methylene groups. These two orbitals are to a large extent (weight of ca. 0.8) linear combinations of the two symmetric delocalized canonical orbitals (see Figure \ref{fig:dmx_canon}). The latter already exhibit almost spin-like parities of ca. -0.9. If those parities were optimal, the ground state of \textbf{dmx} would be a non-disjoint doublet diradical. However, parity optimization results in localization on the different sites, providing evidence that the ground state is rather a disjoint doublet diradical. The lowest triplet state exhibits very similar localized spin-like orbitals as the one with the same  perfect parities, revealing its disjoint character. 
	%
	%
	\begin{figure}[ht]
		\begin{subfigure}[b]{8.3cm}
			\centering
			\includegraphics[width=8.3cm]{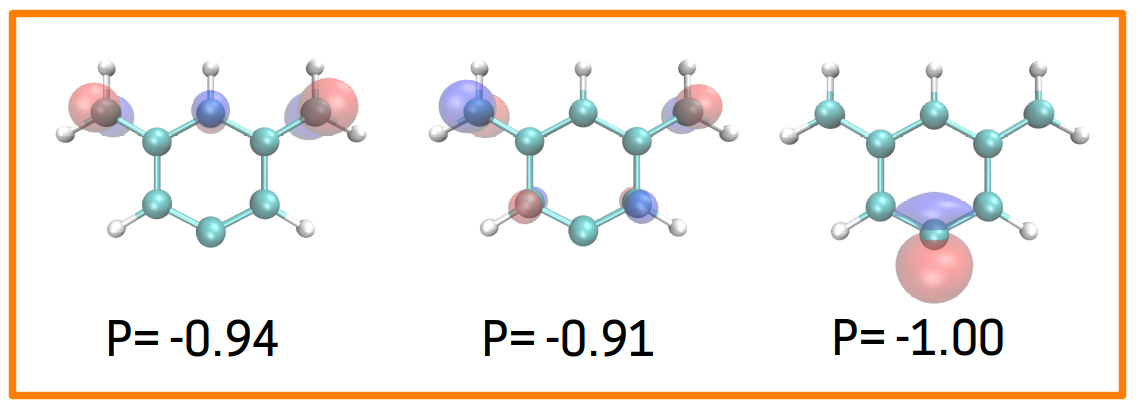}
			\caption{Canonical orbitals}
			\label{fig:dmx_canon}
		\end{subfigure}
		\vfill
		\begin{subfigure}[b]{8.3cm}
			\includegraphics[width=8.3cm]{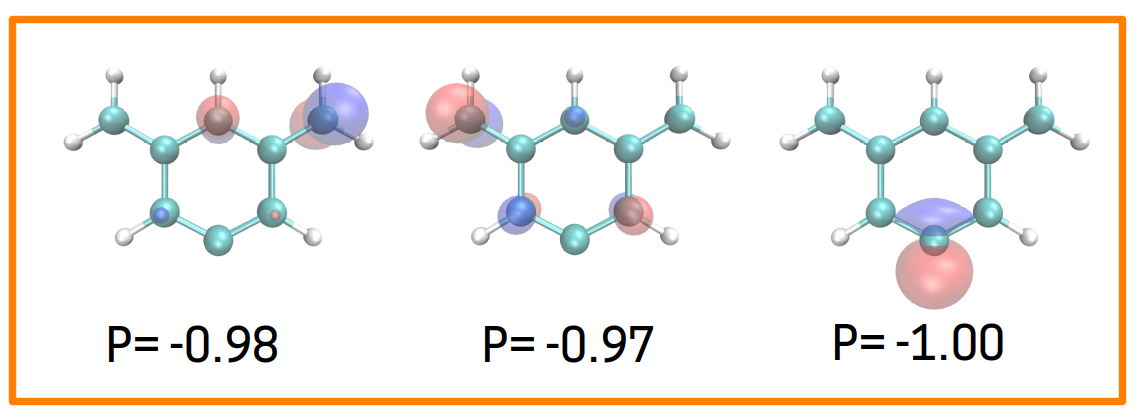}
			\caption{Spin-like orbitals}
			\label{fig:dmx_sm}
		\end{subfigure}
		\caption{Molecular orbitals of the 5-dehydro-m-xylylene (dmx) triradical and their parities: canonical and spin-like.}
		\label{fig:dmx}
	\end{figure}

	An archetypal case of radical emergence is homolytic dissociation of a covalent bonds\cite{helgaker_bible}. Whereas around the equilibrium bond distance a singlet molecule has little if any radical character, the two separated fragments constitute a perfect disjoint diradical with singly-occupied orbitals localized on each side. This is illustrated with our parity-based approach for dilithium (\ce{Li2}) dissociation computed with a simple CASSCF(2,2) wave function (see Figure \ref{fig:li2}). Modestly negative parities of the two active orbitals at around the equilibrium separation (slight diradical character) abruptly fall down in the dissociation region, reaching the limiting value of -2 at bond distance larger than 6 {\AA}. The energy converges to its plateau at approximately the same distance. Interestingly, the minimum parity is achieved for localized 2s-orbitals (obtained from the canonical ones by GVB transformation) at all interatomic distances.
	
	%
	%
	\begin{figure}[ht]
		\includegraphics[width=8.6cm]{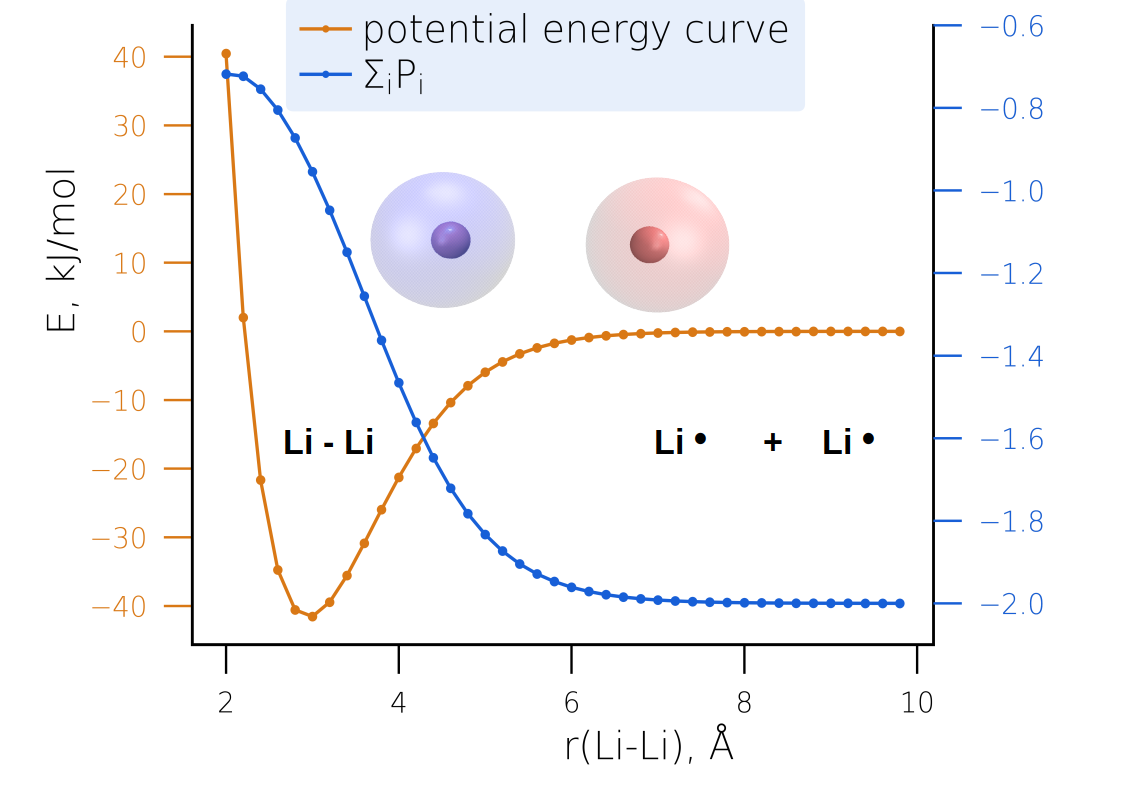}
		\caption{Dissociation of dilithium, \ce{Li2}. Potential energy curve and sum of the two active orbitals' parities for the singlet ground state. Optimal-parity orbitals (centre) at atomic separation of 5 {\AA} are localized on lithium atoms: red - positive; blue - negative.}
		\label{fig:li2}
	\end{figure}

	Next, we consider the evolution of polyradical character in linear oligoacenes with the increase in the number of benzene rings\cite{Bendikov_acenes}. It is instructive to analyze qualitative simple CASCI wave functions in the minimum active space (2,2) including only the HOMO (${\phi}_{1}$) and the LUMO (${\phi}_{2}$) for the two lowest singlet states. These calculations reveal a gradual increase in the measures of radical character for the ground state (see Figure \ref{fig:acenes}) along the series.
	
	For decacene, almost ideal diradical orbital parities of -1 are reached, its wave function being $\Psi_0 = 1/\sqrt{2} (\vert \phi_{1} \bar{\phi}_{1} \rangle  - \vert \bar{\phi}_{2} \phi_{2} \rangle)$, where the bar denotes the opposite spin. The weight of the excited configuration (second term) in the smaller molecules is less than 0.5. The excited state of all homologues is an open-shell singlet and thus a perfect diradical: $\Psi_1 = 1/\sqrt{2}(\vert \phi_{1} \bar{\phi}_{2} \rangle  - \vert \bar{\phi}_{1} \phi_{2} \rangle)$.
	
	Whereas in the excited state delocalized HOMO and LUMO are spin-like, in the ground state minimum parity is achieved for the GVB transformed-orbitals. These orbitals are localized on the opposite fringes of the oligoacene along the short axis (see Figure \ref{fig:acenes}, top). Therefore, the ground state of decacene is a disjoint (zwitterionic) singlet diradical, whereas its excited state (and that of all homologues) is a non-disjoint (covalent) singlet diradical. 
	
	Interestingly, for decacene the 1-RDM, $\mathbf{D}$ for the two states are almost identical, yielding identical odd electron distributions $\mathbf{u}$ (computed as $\mathbf{u} = 2 \mathbf{D} - \mathbf{D}^2$) \cite{Staroverov_2, HEADGORDON2003508} shown in the Appendix. This approach reveals a proper spatial odd electron distribution $u(\mathbf{r})$, but does not provide insights into zwitterionic/covalent characters of diradicals. 
	
	\begin{figure}
		\includegraphics[width=8.6cm]{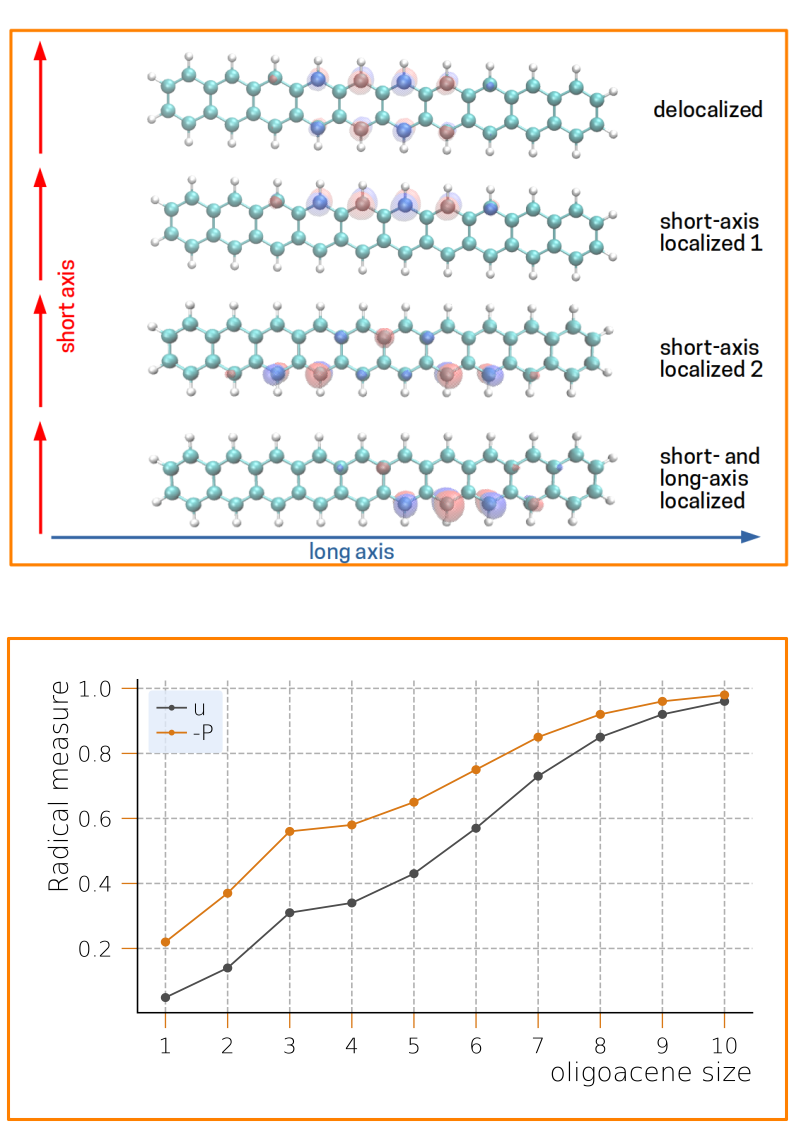}
		\caption{Linear oligoacenes. Top: orbitals of decacene with considerable spin-like character in several electronic states from the CASSCF(4,4) wave function.   Bottom: radical measures (parities and odd electron distributions functions) for the ground state from the CASCI(2,2) wave function, shown for each of the two most spin-like orbitals.}
		\label{fig:acenes}
	\end{figure}
	
	A more realistic state-averaged CASSCF(4,4) calculation of decacene with HOMO-1 and LUMO+1 included in the active space reveals a more complex picture. As expected, the radical character of the frontier-orbital subspace is reduced: the corresponding parities are $-0.85$ for the ground state and $-0.81$ for the previously considered excited state $\Psi_1$. At the same time, near frontier orbitals (HOMO-1 and LUMO-1) contributes to the total radical character: spin-like orbitals in this subspace reach parities of $-0.35$ and $-0.15$ for the ground and excited states, respectively. Remarkably, for the two states the most spin-like orbitals in the near-frontier space are delocalized (short-axis localized) if their counterparts in the frontier space are short-axis localized (delocalized).
	
	The state-averaged CASSCF(4,4) wave function reveals another excited state with the four most spin-like orbitals ($P=-0.65$) localized along both short and long axis (see Figure \ref{fig:acenes}), \textit{i.e.} in the "corners" of the molecule. Those are obtained by mixing all four active orbitals. 
	
	Similar localization patterns and evolving polyradical character in various states of oligoacenes have been shown by Yang \textit{et al.}\cite{Yang_acenes} by manual wave function analysis, whereas spin mapper approach arrives at these results automatically. Despite conceptual simplicity the actual orbital transformation matrices are non-trivial (see the SI) due to the complicated nature of state-averaged CASSCF wave function and can barely be reproduced by hand.

	\section*{Conclusion}
	\label{conclusion}
	We have introduced a theoretical approach to understanding  radicals based on orbital parities, $P$. For single-determinant wave functions $P=-1$ for the singly-occupied and $P=1$ for the doubly-occupied orbitals. Thus, there exist a one-to-one correspondence between occupation numbers and parity. For general multiconfigurational wave functions this is not the case. Therefore, the spin-like orbtials are obtained via the orbital transformation minimizing their parities (spin-mapper transformation) and can be different from canonical and natural orbital. Values of $P$ and the quantity of spin-like orbitals provide qualitative measures of radical character. Visualizing spin-like orbitals proves useful to understand spatial distribution of the unpaired spins, whereas their localization properties reveal the radical type (disjoint/non-disjoint).
	
	We have demonstrated the power of the methodology quantifying radical character of a number of complex di- and polyradicals in several electronic states of different multiplicity, revealing their localization type and visualizing spin-like orbitals. We believe that the orbital-parity approach to radical chemistry, is simple, general, and versatile, being based on fundamental quantities (reduced density matrices) and versatile. It can contribute to the understanding of fundamental radical chemistry by means of theory and provides a useful tool for quantum chemistry practitioners. 
	
	

	

	\section*{Acknowledgements}
	This work was supported by the German Federal Ministry of Economic Affairs and Climate Action through the AQUAS project (Grant No. 01MQ22003A),  German Federal Ministry of Education and Research through the Qsolid project (Grant No. 13N16155) and the Q-Exa project (13N16065).

	\section*{Conflict of Interest}
	
	The authors declare no conflicts of interest.
	
	
\clearpage
	\appendix
	
	\section{Parity optimization}
	\label{seq:opt_parity}
	We apply an iterative procedure to determine the spin-like orbital basis in which the orbital
	parities are extremal.
	For this we perform a sequence of unitary pairwise rotations by an angle $\theta$ of the fermionic operators:
	\begin{align}
	\hat{a}_{q\sigma} &= \cos \theta \, \hat{a}_{i\sigma} + \sin \theta \, \hat{a}_{j\sigma}\,, \nonumber\\
	\hat{a}_{p\sigma} &= -\sin \theta \, \hat{a}_{i\sigma} + \cos \theta \, \hat{a}_{j\sigma}\,,
	\end{align}
	with the same rotation being performed for the hermitian conjugates of the operators.
	From the reduced density matrices we can compute the parity of
	an orbital $\phi_p$, which results from the linear combination of orbitals $\phi_i$ and $\phi_j$, as
	$\langle P_p \rangle_0 (\theta)$.
	The orbital parity $\langle P_{p}\rangle_0(\theta)$ is an analytic, $2\pi$-periodic function of the rotation angle
	$\theta$. We find extremal points $\theta_n$ of the function $\langle P_{p} \rangle_0 (\theta)$ in
	the domain $\theta \in [0, 2 \pi)$ from
	\begin{align}
	\label{eq:cond_rot}
	\left. \frac{d\langle P_{p} \rangle_0}{d\theta}\right\vert_{\theta_n} = 0\,, 
	\end{align}
	and select solutions $\theta_n$ that satisfy
	\begin{align}
	\left . \frac{d^2 \langle P_{p} \rangle_0}{d\theta^2}  \right\vert_{\theta_n} \neq 0 \,.
	\end{align}
	The analytic expression for the derivatives of the function derived and implemented.
	
	%
	%
	\section{Computational details}
	CASSCF \cite{Sun} calculations were carried out with the \textbf{PySCF} package version 2 \cite{Sun1,Sun2,Sun3}. 
	All calculations were done using triple-$\zeta$ Def2-TZVP basis set (with the exception of oligoacenes, for which a smaller def2-SVP basis was used) \cite{Schafer} with the default auxillary density-fitting basis 
	The active space of the CASSCF calculations was comprised of all $\pi$-orbitals and the non-bonding orbitals. The initial orbitals used for CASSCF calculations were UHF natural orbitals \cite{Pulay_Hamilton}.
	The \textbf{Geometric} software package \cite{Wang} has been used for geometry optimization. 
	All the structures except oligoacenes were optimized for the ground-state using the corresponding CASSCF method.
	%
	%

	\newpage
	\section{Odd electron distributions for selected examples}
	
	\begin{figure}[ht]
		\begin{center}
			\includegraphics[width=5.cm]{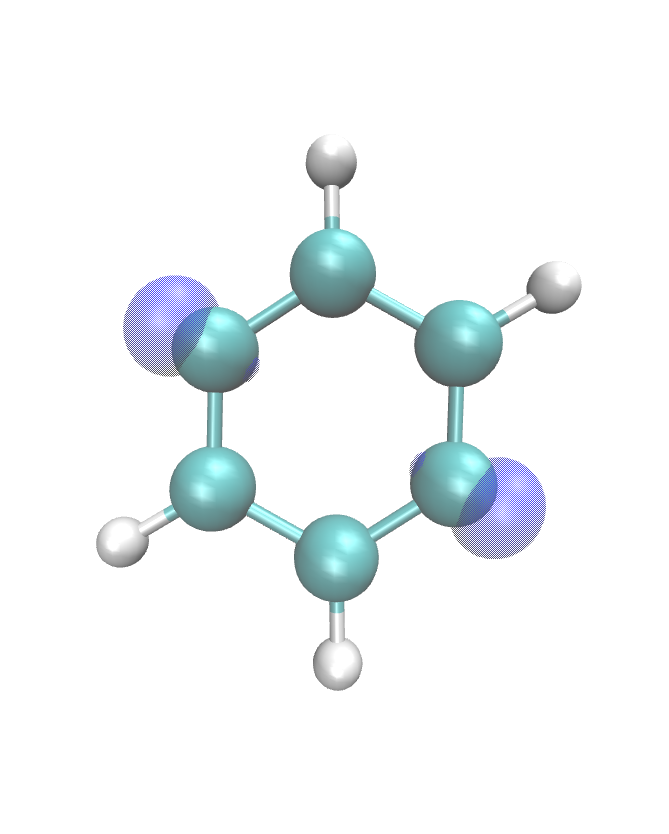}
			\caption{Odd electron distribution for p-benzyne from the CASSCF(8,8) wave function.}
			\label{fig:parities}
		\end{center}
	\end{figure}
	
	\begin{figure}[ht]
		\begin{center}
			\includegraphics[width=8.cm]{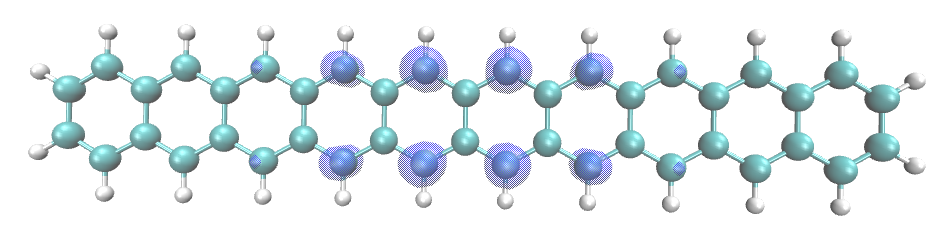}
			\caption{Odd electron distribution decacene from the CASCI(2,2) wave function: the function is identical for the ground and excited singlet states.}
			\label{fig:parities}
		\end{center}
	\end{figure}


	\clearpage
	\bibliography{radicals_refs.bib}

\end{document}